\documentclass[aps, 
amsmath,amssymb, notitlepage, 
nofootinbib,
 twocolumn, 
superscriptaddress,
]{revtex4-1}

\usepackage{graphicx}
\usepackage{color}
\usepackage{bbold}

\newcommand{\bit}{\begin{itemize}}
\newcommand{\eit}{\end{itemize}}

\newcommand{\f}{\frac}
\renewcommand{\>}{\right\rangle}
\newcommand{\<}{\left\langle}
\newcommand{\ba}{\begin{align}}
\newcommand{\ea}{\end{align}}
\newcommand{\be}{\begin{equation}}
\newcommand{\ee}{\end{equation}}
\newcommand{\bi}{\begin{itemize}}
\newcommand{\ei}{\end{itemize}}
\newcommand{\lf}{\left(}
\newcommand{\ri}{\right)}
\newcommand{\dd}{\mathrm{d}}

\graphicspath{{BraidsFigsPRL/}}

\begin{document}

\newcommand{\bra}[1]{\< #1 \right|}
\newcommand{\ket}[1]{\left| #1 \>}

\title{Topological Constraints in Directed Polymer Melts}
\author{Pablo Serna}
\affiliation{Theoretical Physics, Oxford University, 1 Keble Road, Oxford OX1 3NP, United Kingdom}
\affiliation{Departamento de F\'isica -- CIOyN, Universidad de Murcia, Murcia 30.071, Spain}
\author{Guy Bunin}
\author{Adam Nahum}
\affiliation{Department of Physics, Massachusetts Institute of Technology, Cambridge, MA 02139, USA}
\date{\today}

\begin{abstract}
\noindent
Polymers in a melt may be subject to topological constraints, as in the example of unlinked polymer rings. How to do statistical mechanics in the presence of such constraints remains a fundamental open problem. We study the effect of topological constraints on a melt of directed polymers, using simulations of a simple quasi-2D model. We find that fixing the global topology of the melt to be trivial changes the polymer conformations drastically. Polymers of length $L$ wander in the transverse direction only by a distance of order $(\ln L)^\zeta$ with $\zeta \simeq 1.5$. This is strongly suppressed in comparison with the Brownian $L^{1/2}$  scaling which holds in the absence of the topological constraint. It is also much smaller than the predictions of standard heuristic approaches --- in particular the $L^{1/4}$ of a mean-field-like `array of obstacles' model --- so our results present a sharp challenge to theory. Dynamics are also strongly affected by the constraints, and a tagged monomer in an infinite system performs logarithmically slow subdiffusion in the transverse direction. To cast light on the suppression of the strands' wandering, we analyse the topological complexity of subregions of the melt: the complexity is also logarithmically small, and is related to the wandering by a power law. We comment on insights the results give for 3D melts, directed and non-directed.
\end{abstract}

\maketitle

\linespread{1.05}\selectfont

\noindent
The fact that polymer chains cannot pass through each other is the crucial factor in their dynamics, underlying for example the reptation picture \cite{de gennes reptation 1971, DeGennes_book, doi edwards}, and in various situations also determines their equilibrium state. Two salient examples are a single ring polymer and a melt of rings that do not knot or link. In such cases the no-crossing condition sets topological constraints that are inherently non-local. The statistical mechanics of such systems is a tremendous theoretical challenge, for which no systematic theoretical tools are presently available. Heuristic approaches \cite{Cates and  Deutsch, Edwards_AOO, ring_AOO_Rubinstein, ring_AOO_Nechaev,  ring_AOO_dynamics, Grosberg flory arguments, Sakaue_Flory} and ever--growing numerical simulations \cite{muller, suzuki, vettorel, halverson, michieletto, imakaev et al, tamm} have provided substantial insight, but even basic issues --- such as the size of a single ring polymer in a melt or the degree to which different rings mix --- are not resolved. Ring polymer melts have received considerable attention as models of chromosome arrangement in the nucleus \cite{chromosome_review}, and experiments on ring melts have revealed unique rheological properties \cite{melt_viscosity_experiment}. Additionally, dense systems of open chains may be subject to effective topological constraints on intermediate timescales, yielding very slow relaxation and long-lived `pseudoequilibrium' states with  less entanglement than at equilibrium \cite{crumpled globule, fractal globule}.

\begin{figure}[b]
 \begin{center}
 \includegraphics[width=0.98\linewidth]{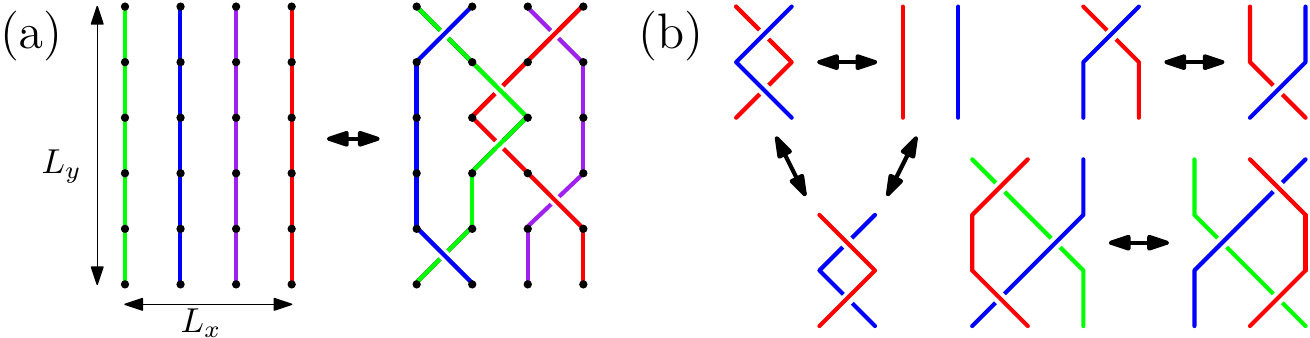}
 \end{center}
 \caption{(a) Topologically trivial configuration in a small system. (b) Monte Carlo move types. There are six variants of the move on the lower right,  and two of that on upper right.}
 \label{method}
\end{figure}

The aim of this paper is to study the simplest possible (but genuinely many-body) model for a topologically constrained melt. Physically, this model describes directed polymers in quasi--2D, i.e. in a slab geometry, but with the positions of the polymers projected onto the plane to give a 2D lattice model. The remnants of three--dimensionality are the fact that the polymers can cross and the crucial distinction between over and undercrossings (Fig.~\ref{method}). The endpoints of the polymers are held fixed, and the entire melt is constrained to be topologically trivial: that is, continuously deformable to the state in which all polymers are straight lines. The melt is endowed with Monte-Carlo dynamics that respect this constraint (i.e., respect the fact that the polymers cannot pass through each other).  Mathematically, the polymers form a trivial `braid'. The statistical properties of random braids have been studied extensively in order to shed light on polymer topology \cite{Nechaev_lecture_notes, Nechaev_long_braid, heap, B3, ferrari review}, but the dynamical and conformational properties of a topologically constrained melt have not been investigated.

Our model is extremely tractable computationally, so we are able to obtain precise results for its universal properties: these turn out to be surprising in the light of current theoretical ideas.  The model is also simple enough to allow hope of analytical progress.

A key feature of the model is that it allows comparison with the predictions of standard theoretical approaches, shedding light on the validity of ideas that are more general than the directed case: for example the idea of modelling topological effects in a melt using toy models of a single polymer in an array of obstacles \cite{ring_AOO_Nechaev, ring_AOO_Rubinstein, ring_AOO_dynamics}, or the use of Flory-like arguments \cite{Cates and Deutsch,Sakaue_Flory, Grosberg flory arguments}. These approaches are widespread but hard to justify a priori. It is important to find ways to confront them with precise results from a genuinely interacting many-polymer model.

The present model may also capture the universal behaviour of some realistic situations. The most striking features of our results are expected to extend to systems of directed polymers in 3D, so the results for dynamical behaviour may be relevant to relaxation and equilibration in polymer brushes \cite{brushes_Alexander, brushes_deGennes}.

In both the present model and 3D ring melts, topological constraints reduce the extension of individual polymers: entropy dictates that polymers `hide' from each other, as more extended configurations are more likely to be entangled. We will soon see  that for directed polymers this effect is almost as strong as it could possibly be.

\begin{figure}[b]
 \begin{center}
 \includegraphics[width=0.95\linewidth]{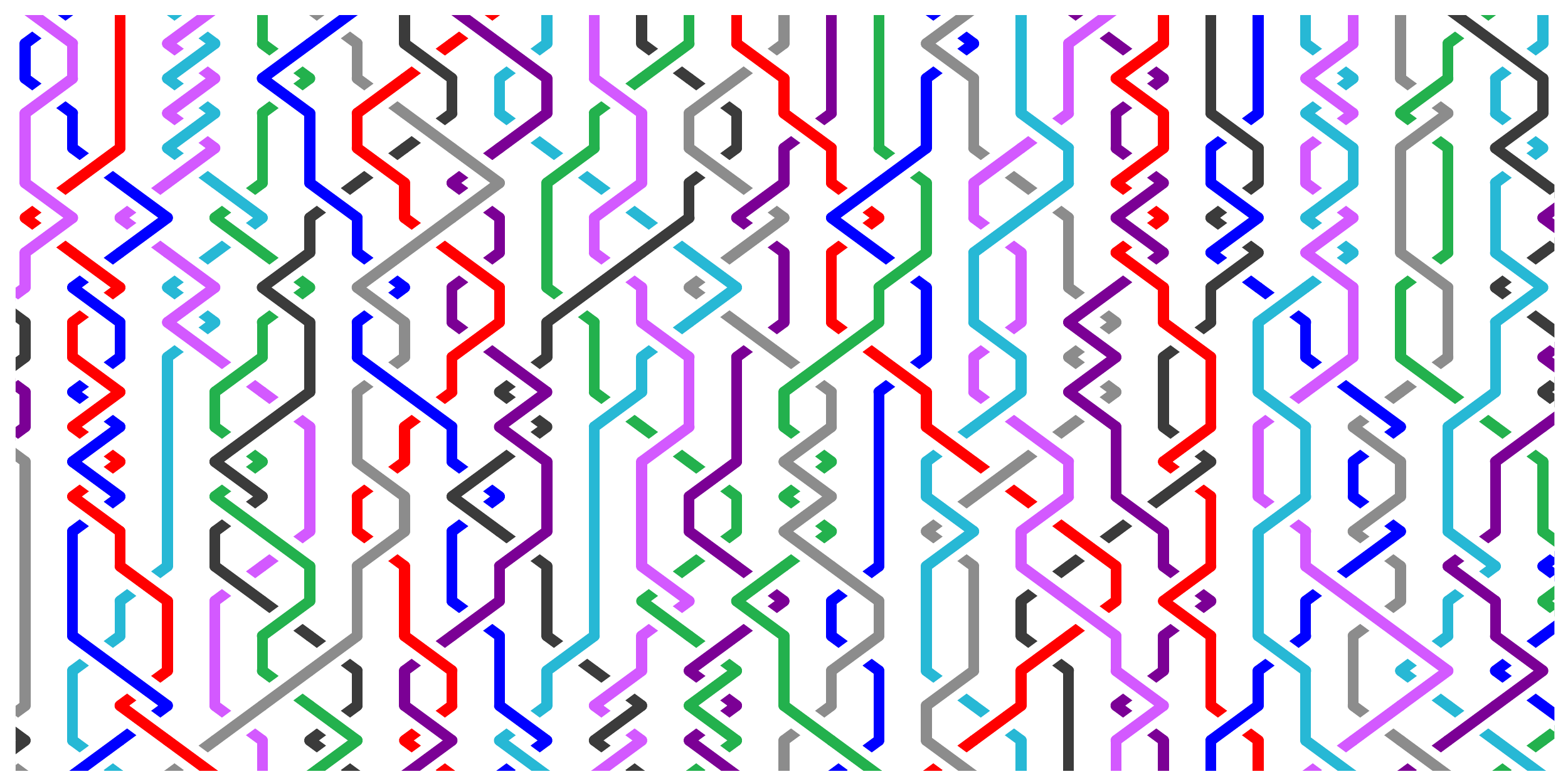}
 \end{center}
 \caption{A subregion of a topologically trivial melt (taken from a system of size $L_x = 512$, $L_y = 128$). }
 \label{meltfigure}
\end{figure}

\begin{figure}[t]
 \begin{center}
 \includegraphics[width=0.92\linewidth]{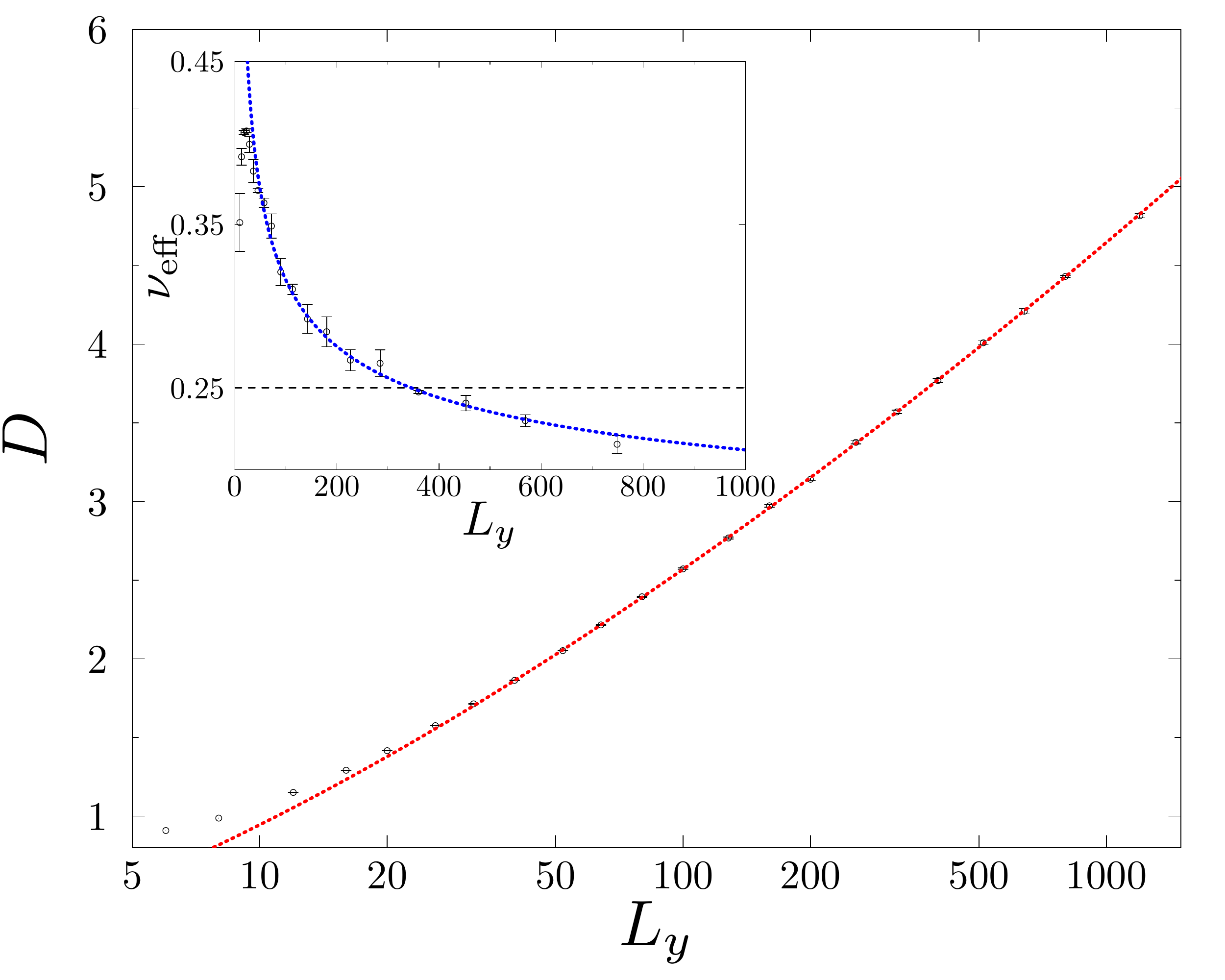}
 \end{center}
 \caption{Main panel: transverse wandering $D$ of a strand at its midpoint $y=L_y/2$, as a function of the length $L_y$ of the strands (note lin-log scale). Red line: fit to the logarithmic form in Eq.~\ref{wandering fit}. Inset: effective exponent $\nu_\text{eff} = \dd \ln D / \dd \ln L_y$, see text, with fit derived from Eq.~\ref{wandering fit}.}
 \label{DvsLy}
\end{figure}

\textit{Model.} Take a number $L_x$ of strands, each of height $L_y$ and directed in the $y$ direction, see Fig.~\ref{method}(a). At integer $y$--values the polymers lie at integer $x$--values, and all lattice points are occupied. Between $y$ and $y+1$, a given polymer may be vertical, or two neighbours may cross. We take periodic boundary conditions in the $x$ direction, and  fix the positions of the endpoints at ${y=0}$ and ${y=L_y}$ so the two ends of a given polymer have the same $x$--coordinate. Finally we enforce topological triviality: the allowed configurations are those which can be deformed to the configuration with straight vertical polymers.  Mathematically, each configuration $C$ defines an element $g(C)$ of the braid group \cite{PBC footnote}, and allowed configurations are those in which this is the trivial element `$1$'. The partition function $Z$ is the equally weighted sum over allowed configurations ($\delta$ is the Kronecker delta):
\be\label{partition function}
Z = \sum_\text{configs $C$} \delta_{g(C),1}
\ee
In practice, to fully define the model we need a Monte Carlo scheme which samples only the topologically trivial configurations. For this we use the moves shown in Fig.~\ref{method}(b), which have a simple relationship with the defining relations of the braid group  \cite{Braids_book}, and which form a complete set of moves. Intuitively, any local rearrangement can be decomposed into: creation/annihilation of pairs of crossings; motion of crossings; and motion of one strand over/under a crossing between two others. These are precisely the moves in Fig.~\ref{method}(b). Fig.~\ref{meltfigure} gives an idea of what a subregion of the melt looks like. Note that the model has a fixed monomer density: this eliminates finite-size effects due to fluctuations of the density mode, which is irrelevant to long-distance behaviour.

We work in the limit where $L_x$, the number of strands, is much greater than the typical wandering of the strands. In fact we enforce a stricter criterion: we ensure $L_x$ is large enough that the results are essentially those of  the $L_x\rightarrow \infty$ limit. We find  that this can be achieved using modest $L_x$, which is unsurprising given that the wandering is much smaller than $L_y$. Below we take $L_y$ ranging up to $L_y = 1200$, and $L_x$ ranging up to $L_x=100$ (larger for some $L_y$). Appendix~\ref{simulations details appendix} gives further details of simulations, including basic checks of equilibration and convergence in $L_x$.

\textit{Results.} The wandering $D_i(y)$ is defined as the transverse displacement of the $i$th strand at height $y$.  We denote the root mean square transverse displacement by  $D(y)$. In the \emph{absence} of the topological constraint in Eq.~\ref{partition function}, each strand essentially performs a random walk constrained to return to $x_{y=0}$ when $y=L_y$: thus it is clear (and we have checked, App.~\ref{unconstrained model appendix}) that in the unconstrained case $D(L_y/2)\sim\sqrt{L_y}$. 

As noted above, the topological constraint will reduce the polymers' wandering. Fig.~\ref{DvsLy} quantifies this using the r.m.s. wandering at the midpoint of the strands, $D(L_y/2)$, plotted against $L_y$. The fit is
\ba\label{wandering fit}
D & = A  \lf \ln \f{L_y}{l_0} \ri^\zeta, & 
\zeta & = 1.49(3) \,,
\end{align}
with $A=0.26(2)$, $l_0=0.91(9)$. 

This logarithmic form is unexpected.  However, a more conventional power law fit $D \propto L_y^{\nu}$ leads to much worse results, or to an exponent equal to zero within error bars if subleading corrections are included \cite{power law footnote}. The inset to Fig.~\ref{DvsLy} shows an effective finite-size wandering exponent defined by $\nu_\text{eff} = \dd \ln D/\dd \ln L_y$. This drifts downwards, as expected for the logarithmic form, according to which $\nu_\text{eff}\rightarrow 0$ as $L_y\rightarrow \infty$. The `array-of-obstacles' prediction discussed below, $\nu = 1/4$, is clearly ruled out. See App.~\ref{simulations details appendix} for further discussion of fits.

\begin{figure}[t]
 \begin{center}
 \includegraphics[width=0.9\linewidth]{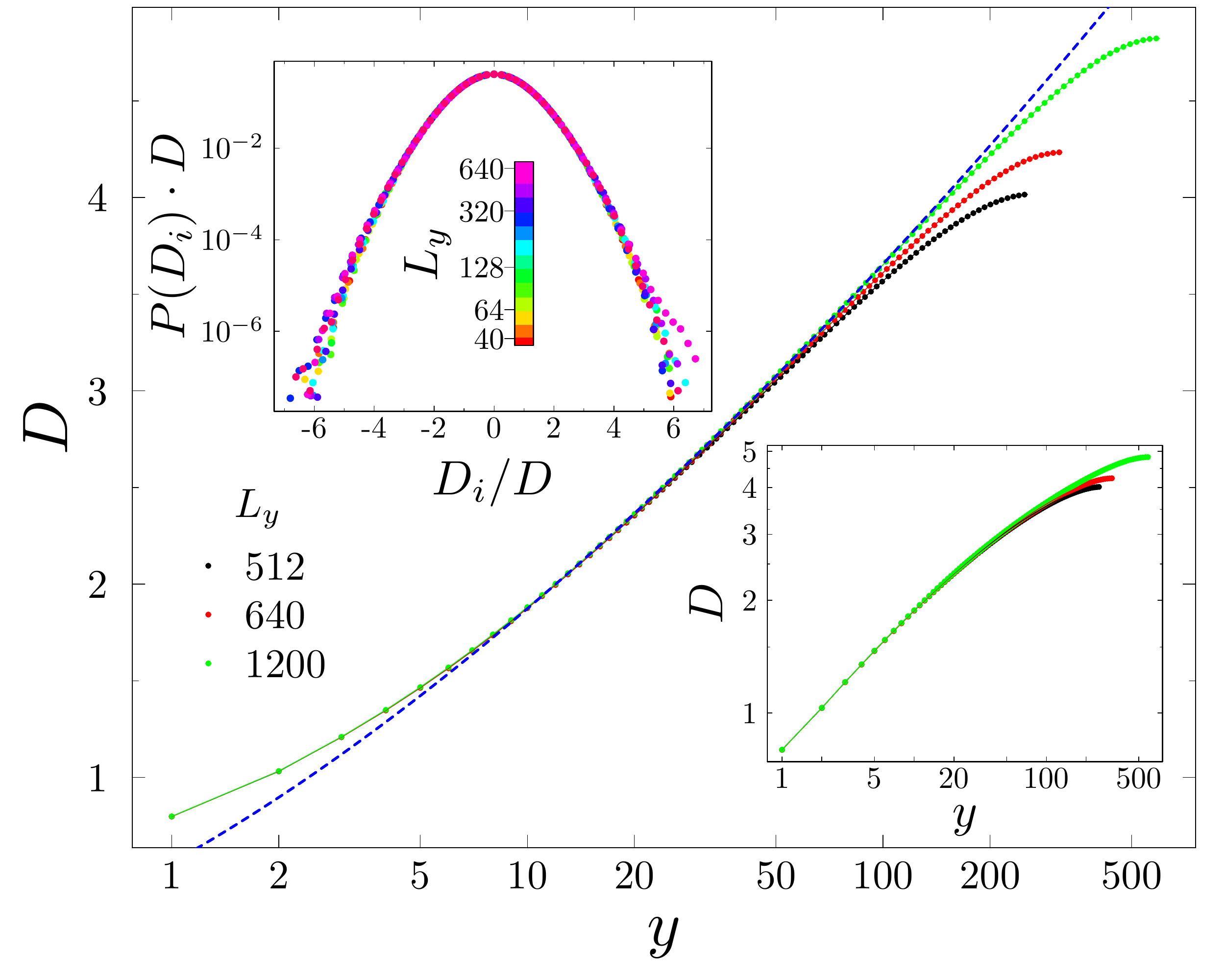}
 \end{center}
 \caption{Main panel: r.m.s. wandering $D(y)$ versus position $y$ along the strand, for various values of the strand length $L_y$. Blue line: fit to $D = A' (\ln y / l_0')^{\zeta'}$ for $10\leq y\leq 100$, giving $\zeta'=1.54(15)$, $A'=0.204(10)$, $l_0'=0.146(10)$. Lower inset: same data on log-log plot. Upper inset: probability distribution for wandering of a strand, $P(D_i)$, rescaled by standard deviation $D$.}
 \label{Dvsy}
\end{figure}

Further evidence for the logarithmic behaviour comes from the wandering $D(y)$ in the distinct regime ${y\ll L_y}$. Since no scaling theory exists for this problem, it is not guaranteed   a priori that the behaviour for $y\ll L_y$ and $y\sim L_y/2$ will be similar, but this turns out to be the case. See Fig.~\ref{Dvsy}, where the data fits well to ${D(y) = A' \lf \ln y / l_0'\ri^{\zeta'}}$ with $\zeta'=1.54(15)$. Note the striking agreement between the independently-determined exponents $\zeta$ and $\zeta'$.  The lower inset to  Fig.~\ref{Dvsy} shows the same data on a log-log scale: it is clear from the curvature that a power law would fit only over a very narrow range of scales. The result for $D(y)$ highlights the fact that the properties of a finite-sized subsystem (of height $y$) are strongly affected by the global topological constraint even in the limit $L_y\rightarrow \infty$.

In addition to the r.m.s. displacement of a strand we may consider the full probability distribution.  Fig.~\ref{Dvsy} (upper inset) shows this for the displacement at $y=L_y/2$. The data collapses beautifully after rescaling by the r.m.s. value. The distribution is not quite Gaussian~(App.~\ref{simulations details appendix}).

\textit{Correlations.} Correlations between the displacements of different strands decay exponentially when their separation is much larger than the wandering. Specifically, let $C_D(x) = \< D_i D_{i+x}\>$, where $D_i$ is the transverse displacement of the $i$th strand at its midpoint. At large $x$,  $C_D(x)\sim e^{-x/\xi}$, with a correlation length $\xi(L_y)$ that grows in a roughly similar manner to $D$ (App.~\ref{simulations details appendix}). The two-point function $C_X(x)$ for the density of crossings (for plaquettes at $y=L_y/2$ separated by a distance $x$) decays exponentially with period 2 oscillations and a correlation length of less than two lattice spacings.

\textit{Dynamics and logarithmic subdiffusion}. The timescale $\tau$ for relaxation of the melt is extracted from the Monte Carlo time series. It is independent of $L_x$ for large $L_x$, but depends nontrivially on the length $L_y$ of the strands:
\ba\label{equilibration time eq}
\tau & \sim L_y^z, &
z & = 2.60(6).
\end{align}
This exponent describes the equilibration of the entire system within the topologically constrained space of states. The transverse motion of a tagged monomer in an infinite system is more interesting. By Eq.~\ref{wandering fit}, we expect that motion of the monomer by a distance $x$ involves rearrangements of segments of height $y\sim \exp(x/A)^{1/\zeta}$, and a time $t$ which is a power law in $y$. This implies that the tagged monomer subdiffuses logarithmically slowly:
\be\label{log subdiffusion eq}
\langle x^2 \rangle \sim \lf \ln t\ri^{2\zeta}.
\ee
Similarly, the decorrelation time of a subregion of the melt of size $x\times y$ is exponentially large in $x$.

Eq.~\ref{log subdiffusion eq} describes a monomer inside the topologically trivial melt. It is  interesting to ask about the dynamics of a monomer for other choices of the initial state; for example an equilibrated state of the topologically \textit{un}constrained problem. A naive guess might be that the increased local entanglement in such a state will slow the motion even further, but this needs investigation. These issues are related to relaxation in polymer brushes \cite{reith brush dynamics}, in which the polymers are tethered at one end and are directed (on large scales) for high surface fraction \cite{brushes_Alexander, brushes_deGennes}.

\begin{figure}[t]
 \begin{center}
\includegraphics[width=0.85\linewidth]{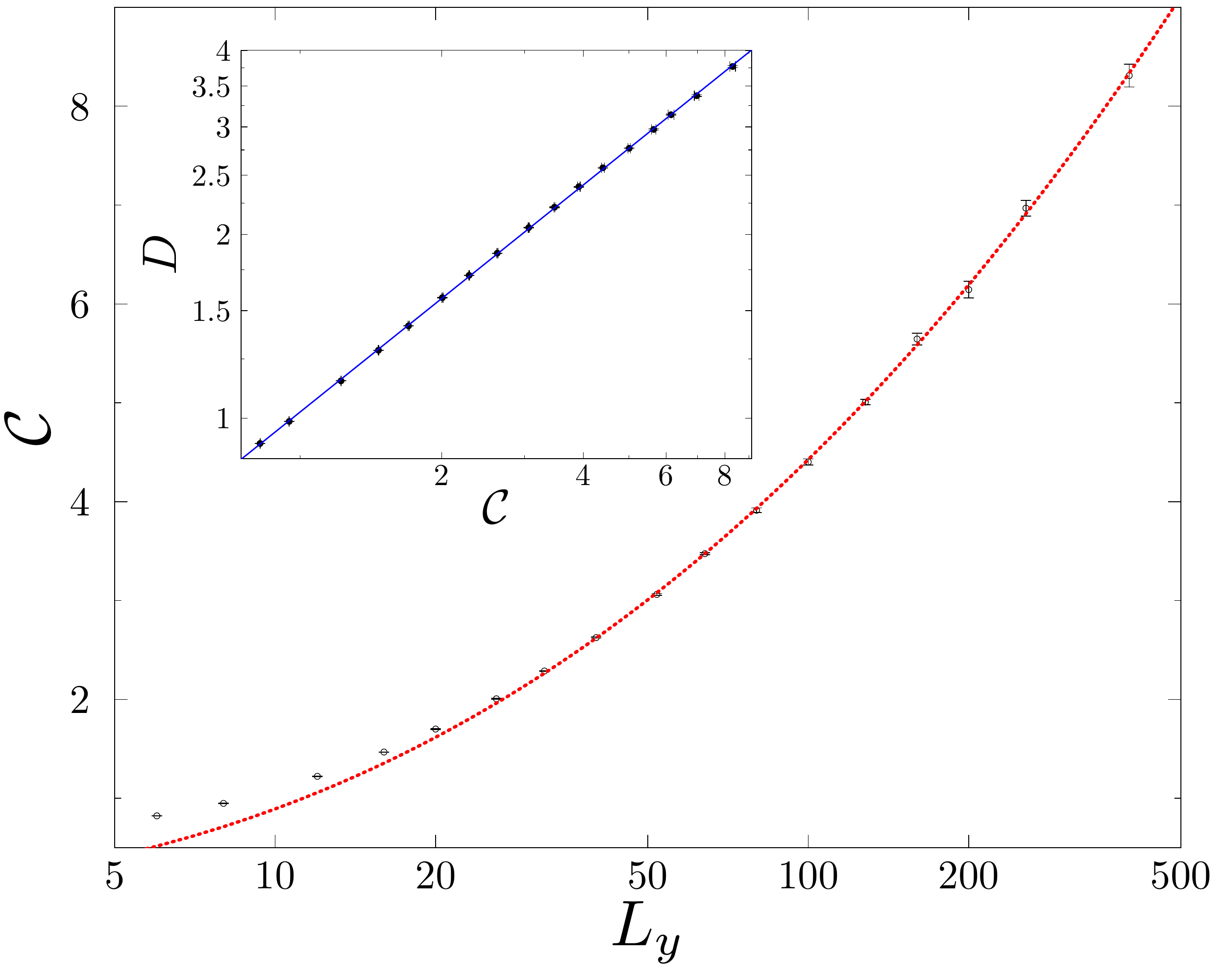}
 \end{center}
 \caption{Main panel: complexity (per strand) for the half-braid as a function of $L_y$, fitted to the form in Eq.~\ref{complexity scaling} with $\eta = 2.6(2)$. Inset: Wandering $D(L_y/2)$ plotted versus the complexity $\mathcal{C}(L_y/2)$ and fitted to a power law, Eq.~\ref{wandering against complexity}.}
 \label{Complexity}
\end{figure}

\textit{Topological complexity.} The present model allows for a clean definition of the topological complexity of a subregion (the full system is of course topologically trivial). Examining this complexity illuminates the drastic suppression of the wandering. We take the subregion to be the bottom half of the braid, $y<L_y/2$.  Fixing the strands' endpoints at $y=0$ and $y=L_y/2$ gives the half-braid a well-defined topology which cannot be changed by allowed moves in the interior. Allowed moves can however reduce the total number of crossings. Let $N_\text{min}$ be the \emph{minimal} value to which we can reduce this number, and define the complexity per strand as
\be
\mathcal{C} =2 N_\text{min}/{L_x}.
\ee
$\mathcal{C}$ is the average number of crossings encountered (i.e. steps to the right or left taken) by a  \emph{single} strand \textit{in the reduced configuration}. $\mathcal{C}$ is finite as $L_x\rightarrow\infty$.  Note that reducing the half-braid does not change $D_i(L_y/2)$: strands in the reduced half-braid take many fewer steps, but wander by the same total distance.

We compute $\mathcal{C}$ by simulated annealing. Starting with an equilibrated braid, we extract the lower half and subject it to a modified Monte Carlo dynamics with an energy penalty for crossings. The temperature is gradually lowered until the system finds its `ground state'. We do not encounter problems finding the ground state (App.~\ref{simulations details appendix}), perhaps because $\mathcal{C}$ is modest.

The main panel of Fig.~\ref{Complexity} shows $\mathcal{C}$ plotted against $L_y$, and the inset shows the r.m.s. wandering $D(L_y/2)$ plotted against $\mathcal{C}$. Strikingly, the wandering has a clean power law dependence on the topological complexity:
\ba\label{wandering against complexity}
D & \propto \mathcal{C}^{\alpha}, & 
\alpha & = 0.618(2).
\end{align}
This implies, for consistency with Eq.~\ref{wandering fit}, 
\ba\label{complexity scaling}
\mathcal{C} & \propto  \lf \ln \f{L_y}{l_0}\ri^{\eta},  & 
\eta & = \f{\zeta}{\alpha} \simeq 2.41(6).
\end{align}
This is indeed compatible with the results in Fig.~\ref{Complexity}. For comparison, a braid configuration of height $L_y/2$ chosen uniformly from the set of all such braids has a typical complexity of order $L_y$ \cite{Nechaev_long_braid, heap}.  For a fixed \emph{finite} number of strands, a sub-braid of sufficiently large height $y$ is believed to have a complexity of order $\sqrt{y}$ \cite{Nechaev_long_braid, Nechaev_lecture_notes, imakaev et al}, but this is a different (`quasi--1D') limit (App.~\ref{simulations details appendix}).

The suppression of wandering may therefore be viewed as a consequence of the drastic suppression of complexity. Interestingly, the geometry of the strands \emph{in the reduced half-braid} is more conventional than in the unreduced half-braid: the wandering $D$ of a strand has a power law dependence on the average number of steps, $\mathcal{C}$. The exponent $\alpha$ in Eq.~\ref{wandering against complexity} is greater than 1/2, indicating positive correlations between steps in the reduced half-braid. By contrast, the number of steps per strand in the \emph{unreduced} half-braid is much larger, $O(L_y)$, and there are strong negative correlations between steps.

\textit{Comparison with standard ideas.} Many approaches to topologically constrained melts rely either on simplifying the problem to a single strand in an array of obstacles which represent the other polymers \cite{Edwards_AOO, ring_AOO_Nechaev, ring_AOO_Rubinstein, ring_AOO_dynamics, rosa everaers}, or on Flory-like free energy arguments \cite{Cates and Deutsch,Sakaue_Flory, Grosberg flory arguments}. These ideas have for example been used to argue that rings in a 3D melt fold up into compact tree-like structures \cite{Grosberg flory arguments}. However both approaches are uncontrolled approximations which must be tested against data. For 3D ring melts this is challenging because of large finite-size effects \cite{rosa everaers, halverson}. 

Here, we can make a quantitative comparison with the natural array of obstacles model for the directed case, which describes a single fluctuating strand in an array of straight vertical strands (App.~\ref{heuristic arguments appendix}). This predicts $D\sim L_y^{\nu}$ with $\nu =1/4$, and $\mathcal{C}\sim L_y^{1/2}$ \cite{ring_AOO_Nechaev}, contrary to our results. The wandering distribution also differs (App.~\ref{simulations details appendix}). The exponent $z$  on the other hand  is roughly compatible with the $z=5/2$ of the array of obstacles model \cite{ring_AOO_dynamics} (though, by the reasoning preceding Eq.~\ref{log subdiffusion eq}, the transverse diffusion in the topologically constrained ensemble will be much faster for the array of obstacles). Our results show that, for directed polymers, the behaviour of the true melt is very different from the array of obstacles model.

Flory-like estimates are sensitive to the assumed form of the entropic cost of not being entangled \cite{Cates and Deutsch, Grosberg flory arguments, Sakaue_Flory}, which is hard to control. Here one can easily show that 2--strand interactions alone are not enough to explain $\nu=0$. See App.~\ref{heuristic arguments appendix} for further discussion. The approach does however support the expectation that wandering is at least as strongly suppressed in 2+1D as in 1+1D.

\textit{Throwing out 3-strand moves.} Though much simpler than a melt of 3D rings, the present model is still a formidable challenge analytically. One may also consider a reduced model based on the `locally free group' \cite{Nechaev_long_braid, heap}, a simplification of the braid group. This means imposing the stronger constraint that  the melt  be deformable to the straight-line state \emph{without} using the 3-strand moves of Fig.~\ref{method}. This is a drastic simplification,  and no longer faithful to the topology of directed melts. Nevertheless preliminary simulations  suggest that behaviour for $D$ remains qualitatively similar, with a reduced ${\zeta\sim 1}$~(App.~\ref{locally free appendix}).

\textit{Future directions.} We believe that the crucial features of the present model, including the fact that the wandering is logarithmic (though not necessarily the value of $\zeta$) will carry over to the 3D directed case. This is because the number of other strands encountered by a given strand grows faster with $D$ in 3D than in 2D, indicating a stronger entropic penalty for wandering. This conjecture must be examined numerically. Another natural next step is to investigate the dynamics of the present model when the endpoints of the chains free to move, so that the topology of the melt can slowly relax. It would be interesting to know whether the transverse motion of the monomers remains logarithmically slow even in the final equilibrated state. If we start from an unentangled configuration, even the static properties may remain similar to those discussed here for a very long time. 

We have seen that for a topologically constrained ensemble of directed polymers, the exponent governing the chains' extension takes its minimal possible value, $\nu = 0$, with logarithmic corrections. One might wonder whether in a 3D ring melt the radius of gyration is also governed by logarithmic corrections to the minimal exponent value ($\nu=1/3$). Conceivably, such logarithms might partially explain the slow saturation observed for $\nu$ in this case.

A fundamental question is whether there exists a real-space renormalisation group treatment for topologically constrained polymers (App.~\ref{heuristic arguments appendix}). The present model may be simple enough to offer hope of this.

\textit{Acknowledgements.} It is a pleasure to thank J. Chalker, J. Haah, M. Kardar and S. Nechaev for useful discussions and comments. This work was supported in part by Spanish MINECO and FEDER (UE) grant no. FIS2012-38206 and MECD FPU grant no. AP2009-0668. PS acknowledges the support of EPSRC Grant No. EP/I032487/1. GB acknowledges the support of the Pappalardo fellowship in Physics. AN acknowledges the support of a fellowship from the Gordon and Betty Moore Foundation under the EPiQS initiative (Grant No. GBMF4303).

\appendix

\section{Further details of simulations}
\label{simulations details appendix}

\noindent
The appendices include: further details of simulations, fits etc. (this section); a discussion of the effect of 3-strand moves (App.~\ref{locally free appendix});  comparison with the topologically unconstrained case (App.~\ref{unconstrained model appendix}); and further information about the heuristic approaches mentioned in the main text (App.~\ref{heuristic arguments appendix}).

\begin{figure}[t]
 \begin{center}
 \includegraphics[width=0.9\linewidth]{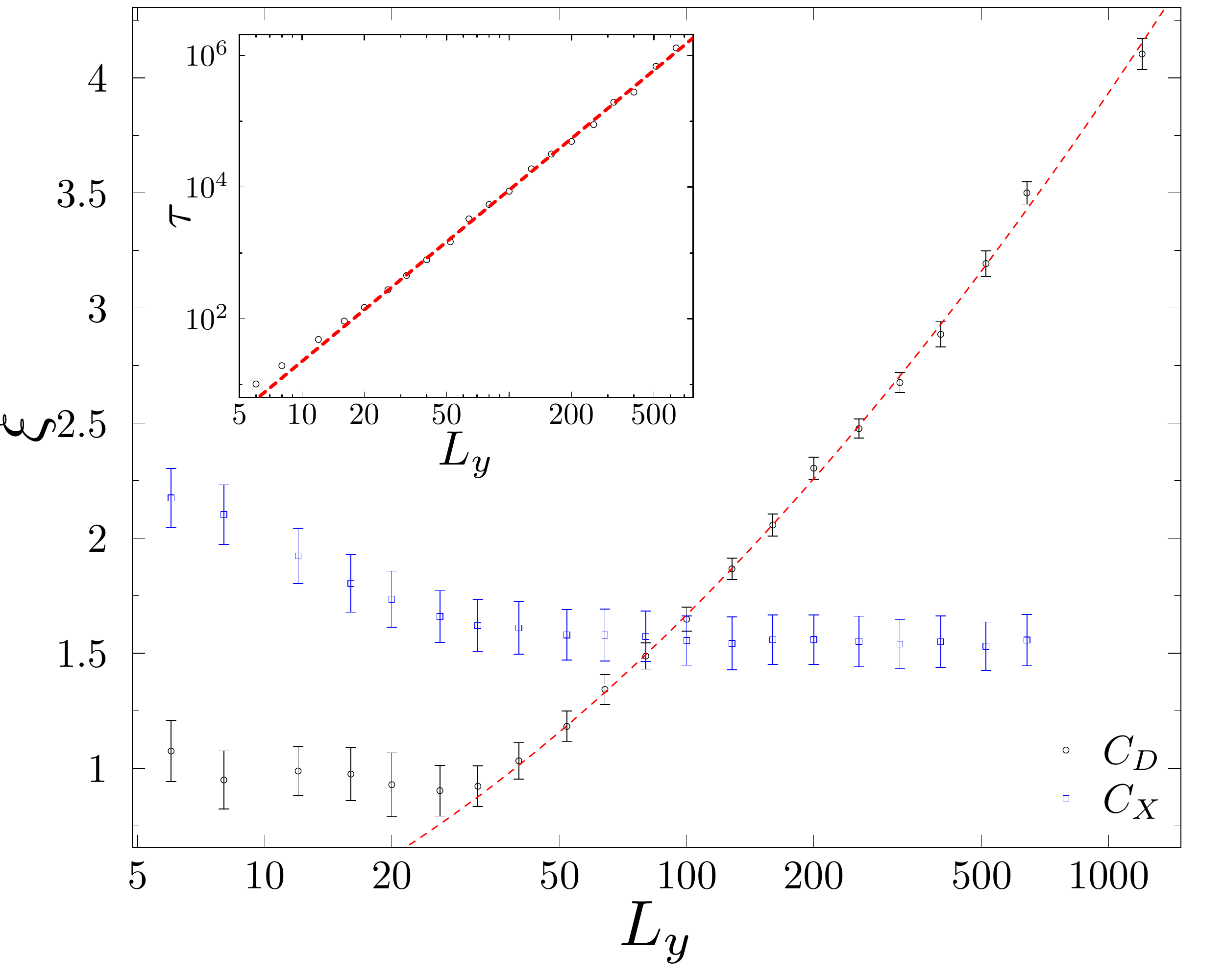}
 \end{center}
 \caption{Main panel: correlation lengths $\xi$ for (1) correlation $C_D$ between the wandering of two strands, and (2) correlator $C_X$ of the density of crossings. The fit (red line) for (1) is to $\xi_D=a (\ln L_y/l_0)^b$ with $a = 0.14(8)$, $l_0=2.1(11)$, $b=1.8(2)$. Inset: correlation time as a function of $L_y$, fitted to $\tau \propto L_y^z$ with $z = 2.60(6)$.}
 \label{correlation time}
\end{figure}

\textit{Equilibration.} An important step is to ensure the equilibration of the sample. To begin with, we calculate a relaxation time $\tau(L_x, L_y)$ from the time series of the r.m.s. wandering $D$ in a standard way, by fitting the autocorrelation function to an exponential. For the $L_y$ values considered, $\tau(L_x,L_y)$ rapidly approaches a constant independent of $L_x$ (on a scale much less than $100$). The $L_y$--dependence of the autocorrelation time is plotted in the inset to Fig.~\ref{correlation time}.

\begin{figure}[b]
 \begin{center}
  \includegraphics[width=0.9\linewidth]{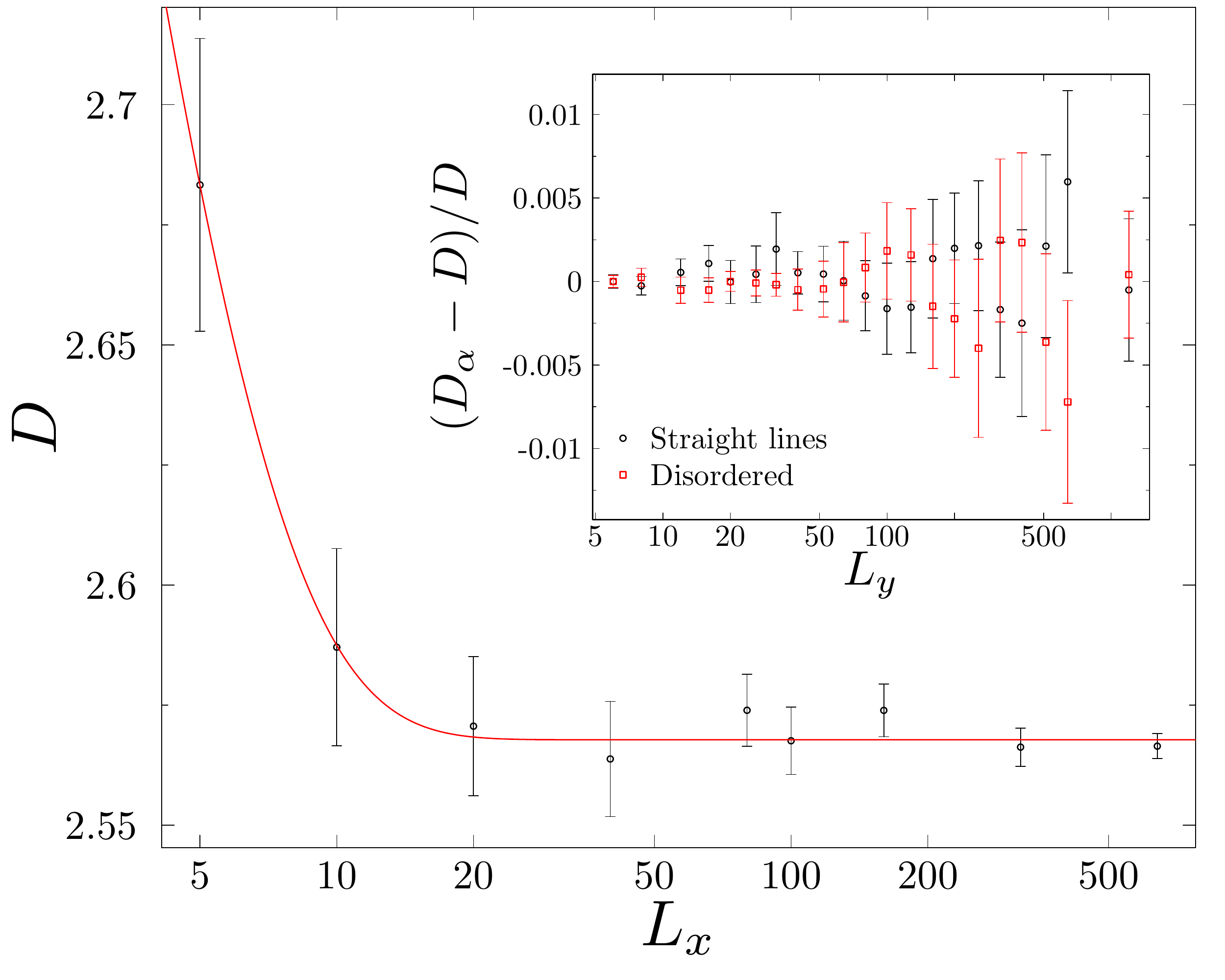}
 \end{center}
 \caption{Main Panel: Wandering $D$ as a function of $L_x$ for samples of size $L_x\times100$. Inset: Relative error in wandering for the two different initial configurations, $(D_\alpha-D)/D$, as a function of $L_y$ ($\alpha = 1,2$).}
 \label{comparis}
\end{figure}

For a stringent check on the equilibration of our samples, at each system size we run simulations starting from two very different  (topologically trivial) initial states. The first is the configuration where all polymers are straight vertical lines. The second is constructed by generating a lower half with crossings at random and taking the upper half to be the mirror image of the lower one. This gives a configuration which is globally trivial but locally highly entangled. The two cases represent opposite extremes both for the wandering at $y=L_y/2$ (which is exactly zero in the first case and $O(L_y^{1/2})$ in the second) and for the half-braid complexity $\mathcal{C}$ (zero and $O(L_y)$ respectively). Nevertheless they converge to the same equilibrated value of the wandering under the Monte Carlo dynamics.  After 20 autocorrelation times, results from straight lines and maximally disordered samples are identical to within error bars: the inset to Fig.~\ref{comparis} shows $(D_\alpha-D)/D$, where $D_\alpha$ is the wandering for samples with different initial configurations $\alpha=1,2$ (straight lines or disordered) and $D$ is the average of both. We therefore start collecting data after $20\tau$ Monte Carlo sweeps.

\textit{Convergence in $L_x$.} We examined the dependence of $D$ and other quantities on the horizontal system size $L_x$ in order to ensure that the results are those of the large $L_x$ limit.  We studied the nature of this convergence extensively for $L_y=6$, $32$ and $100$. We see exponential convergence in $L_x$ with a very small characteristic length and  signs of oscillations within the envelope. The main panel of Fig.~\ref{comparis} shows the wandering for $L_y=100$ plotted as a function of $L_x$: the red line is an exponential fit with a characteristic length $l_0=2.8\pm1.1$. For larger $L_y$ values, we checked that varying $L_x$ did not change the value of $D$, within error bars. (For most samples we considered $L_x \leq 100$, and for $L_y=800$ we considered $L_x\leq 200$.)

Correlation functions also yield short correlation lengths, as discussed in the text and shown in Fig.~\ref{correlation time}.

\textit{Non-Gaussianity of $P(D_i)$.} The probability distribution $P(D_i)$ in Fig.~4 of the main text is close to, but measurably different from, a Gaussian. To see this we examine the difference of moment ratios ${\mathcal{M}_k=\< |D_i|^k\>/\<D_i^2\>^{k/2}}$ from the Gaussian value: ${\widetilde{\mathcal{M}}_k=\mathcal{M}_k-\mathcal{M}_k^\text{Gaussian}}$. For comparison, we have also obtained these moments for the toy model of a single strand in an array of obstacles (see below), by a separate simulation.

For $\widetilde{\mathcal{M}}_{3/2}$, an extrapolation to $L_y=\infty$ gives $-0.021(1)$ for the full model and $-0.0363(3)$ for the array of obstacles, indicating that the universal scaling function is different in the two cases. The error bars are statistical errors in extrapolations of the form $\widetilde{\mathcal{M}}=A + B L_y^{-c}$; systematic errors may be larger. For the fourth cumulant the extrapolation to $L_y=\infty$ is more difficult but we find $\widetilde{\mathcal{M}}_4 \sim 0.2$ for the full model and $\widetilde{\mathcal{M}}_4\sim 0.5$ for the toy model.

\begin{figure}
 \begin{center}
  \includegraphics[width=0.8\linewidth]{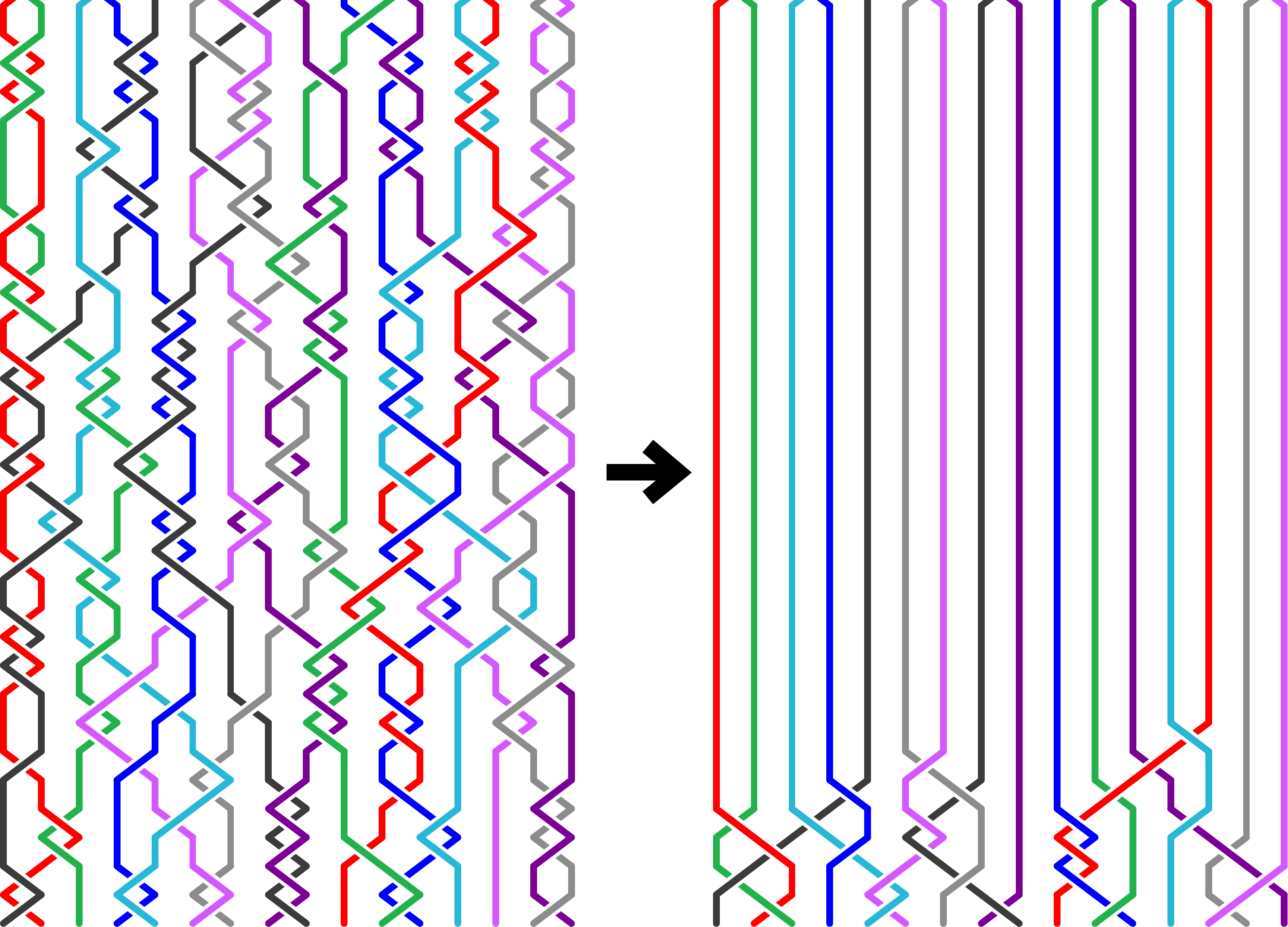}
 \end{center}
 \caption{The reduction of a half-braid. (For illustrative purposes only:  results in  text use larger $L_x $ values.)}
 \label{reductionfig}
\end{figure}

\textit{Computation of topological complexity of half-braid.} To check the simulated annealing protocol described in the text, we reduce the same half-braid multiple times and confirm that after each iteration the same number of crossings remains. We also check that the protocol succeeds in reducing the full braid to the straight-line configuration. Finally we check that reducing the rate at which the temperature is decreased does not change the results. For illustrative purposes, Fig.~\ref{reductionfig} shows a small system before and after reduction.

It has been conjectured that  for a fixed \emph{finite} number of strands, in the limit of large $L_y$, subregions of height $y$ of a trivial braid have a complexity proportional to $\sqrt{y}$ for sufficiently large $y$ \cite{Nechaev_long_braid,Nechaev_lecture_notes,imakaev et al}. Our result that $\mathcal{C}$ is only logarithmically large is surprising in the light of this, but the  results are not in contradiction: the former is for the `quasi-1D' situation, in which $L_x$ is fixed as $L_y\rightarrow \infty$, whereas we consider the 2D situation in which $L_x$ is much larger than the scale of the wandering of the strands. It would be interesting to understand the crossover between the two limits.

\textit{Further checks on logarithmic fits}. In the main text we fitted the wandering (Figs.~3,~4 of the main text) to the forms $D(L_y/2) = A (\ln L_y/l_0)^\zeta$ and $D(y) = A' (\ln y/l_0')^{\zeta'}$, finding remarkable agreement between the exponents $\zeta$ and $\zeta'$. Here we check the extent to which finite-size corrections to these forms could affect the values of $\zeta$ and $\zeta'$. (We have already noted in the text that power-law fits are much less convincing than logarithmic ones.) 

We have tried various possible forms for subleading corrections. Fits with $\zeta=1$, such as ${A \ln(L_y/l_0) +B / (\ln L_y/l_0)}$ or $A \ln (L_y/l_0) + B / L_y$, are poor. For a more stringent test we consider
\be\label{four parameter fit}
D(L_y/2) = A 
\lf \ln \f{L_y}{l_0} \ri^\zeta 
\lf 1 + \f{B}{\ln L_y/l_0}  \ri
\ee
and the analogue for $D(y)$. The resulting exponent values are $\zeta=1.49(7)$ and $\zeta'=1.59(22)$, where the error bars are calculated using the bootstrap method. For $\zeta$, the range used for the fit is $L_y\geq 40$ (we have checked that the exponent is stable when this value is increased) and for $\zeta'$, where we must ensure $y\ll L_y$, it is $10\leq y\leq 100$.

The fact that the values of $\zeta$ and $\zeta'$ are stable under the addition of subleading terms to the fit, and in particular the agreement between $\zeta$ and $\zeta'$, gives us confidence that these exponents are indeed the same, and that they are close to $1.5$ rather than being equal to any integer value.

\begin{figure}[t]
 \begin{center}
\includegraphics[width=0.9\linewidth]{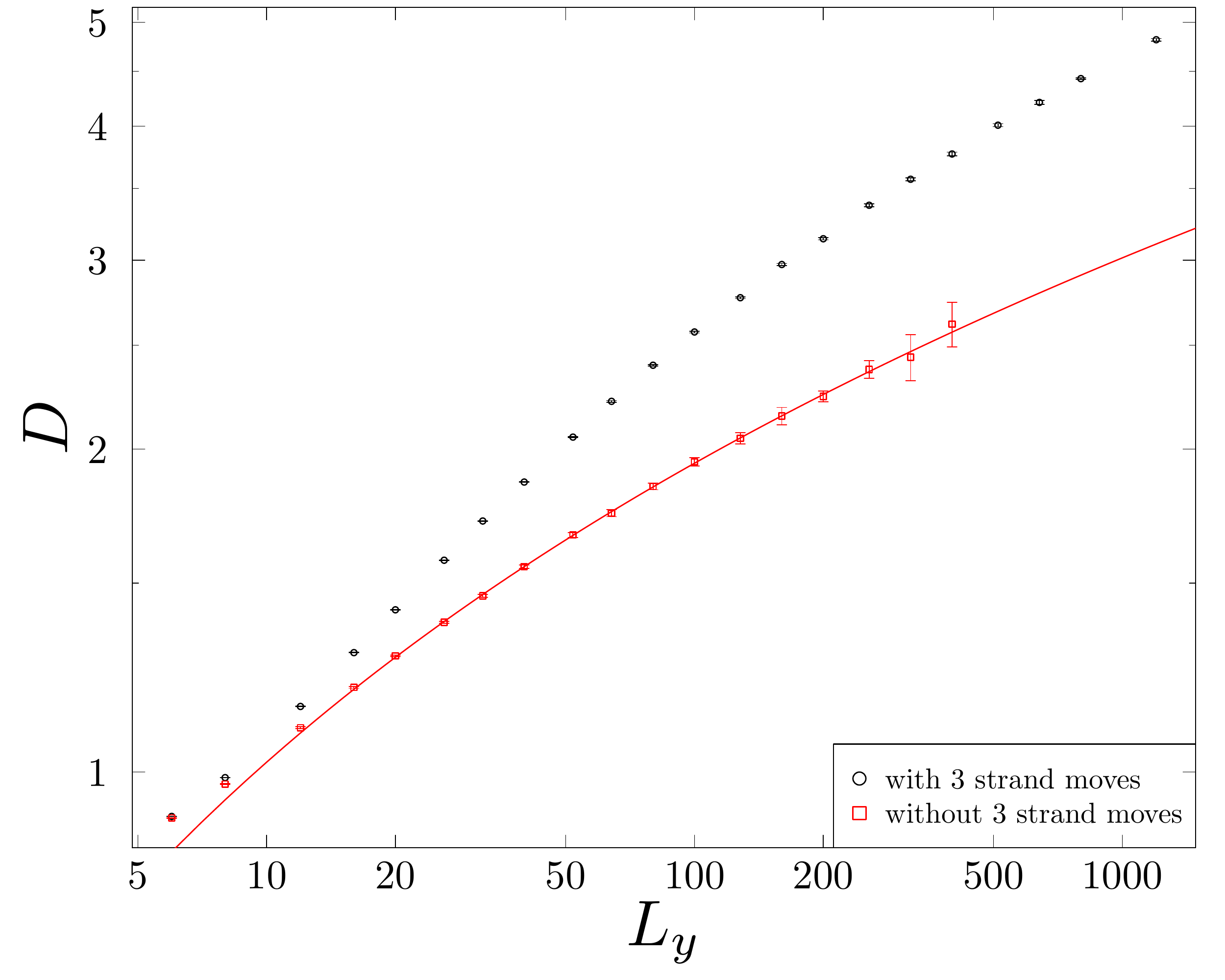}
 \end{center}
 \caption{Comparison of the wandering $D(L_y/2)$ for the full model (same data as in the main text, black) and the model without the three-strand moves (red). Fit described in text.}
 \label{nothreestrandmovefig}
\end{figure}

\section{Simulations without 3-strand moves}
\label{locally free appendix}

\noindent
We have performed limited simulations without using the 3-strand moves of Fig.~1 of the main text, starting with the vertical line initial condition. This corresponds to a partition function in which the polymer configurations satisfy a stronger constraint. This constraint does not have as natural an interpretation in terms of topology of strands: it means that strands are forbidden from moving over crossings between other pairs. However it has been suggested as a natural simplification of the algebraic structure of the problem, equivalent to replacing the braid group with the locally free group \cite{Nechaev_long_braid, heap}.

Fig.~\ref{nothreestrandmovefig} compares $D(L_y/2)$ for the full model discussed in the main text with that for the model without 3-strand moves. A fit to the form $A'' (\ln L_y/l_0'')^{\zeta''}$ for the latter gives $A''= 0.5(3)$, $l_0''=0.9(1.1)$, and $\zeta''=1.0(3)$.

\section{Unconstrained model}
\label{unconstrained model appendix}

\begin{figure}
 \begin{center}
 \includegraphics[width=0.9\linewidth]{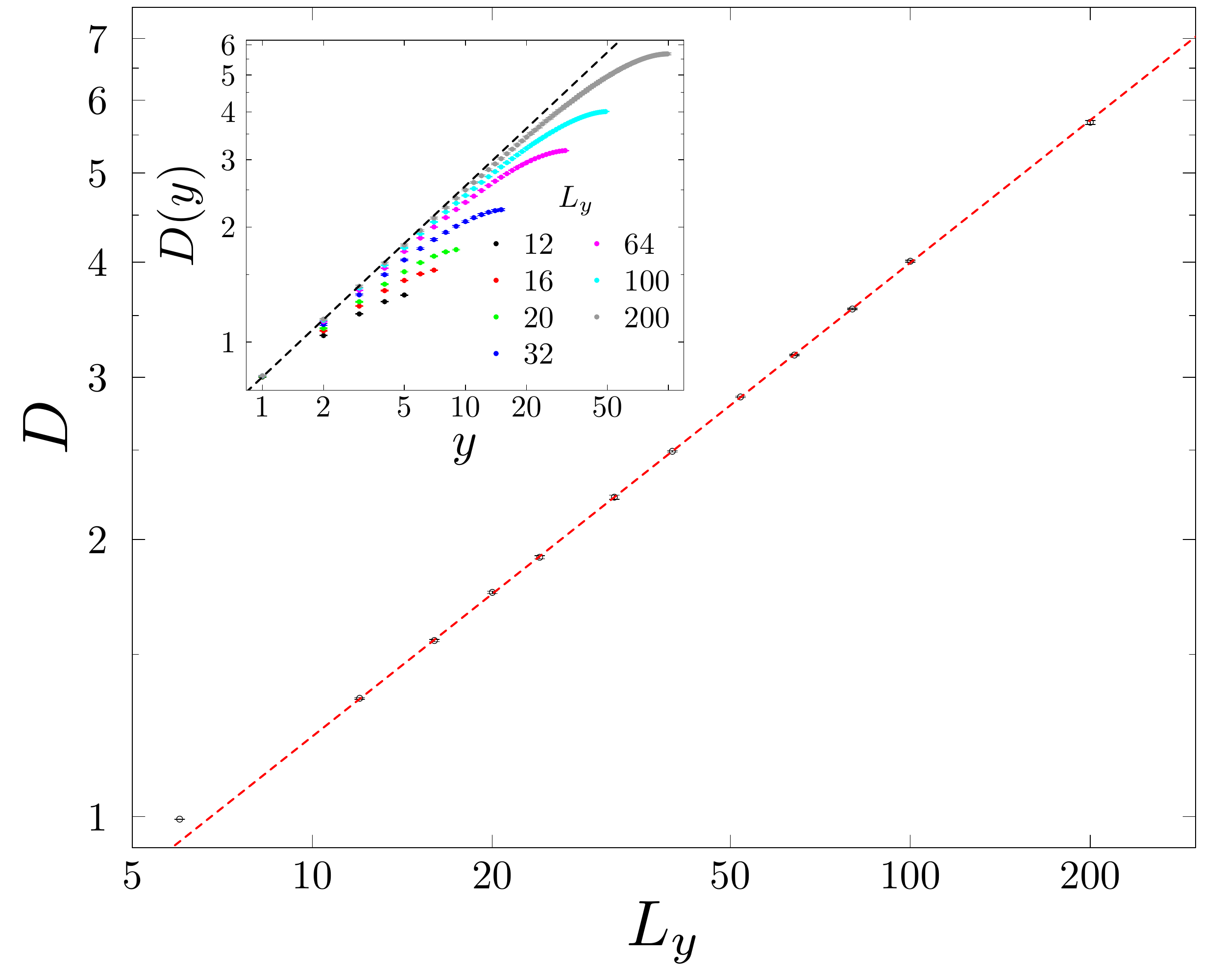}
 \end{center}
 \caption{Main Panel: Wandering as a function of system size in the topologically unconstrained problem, fitted to a power-law with exponent 0.514(2). Inset: Wandering as a function of $y$ for several system sizes: straight line is a power law with Brownian exponent $0.5$.}
 \label{Unc wandering}
\end{figure}

\noindent
As basic check of our algorithms, and to confirm that the unusual features found in the text (e.g. the drastic suppression of wandering) are due to the topological constraint, we have characterised the model without this constraint. As expected it shows simple Brownian wandering as a function of the height $y$ (not to be confused with dynamics in time $t$), and exponentially decaying correlation functions. 

To do this we allow an additional move that exchanges undercrossings and overcrossings. If we also relax the boundary conditions at the top of the sample (so that a strand is allowed to have different $x$--coordinates at $y=0$ and $y=L_y$), the model becomes easily solvable analytically, giving a useful check on our algorithms.

We also checked the case where the topology-changing moves are allowed but the endpoints are kept fixed. The corresponding partition function  is a sum over configurations of any topology but with fixed endpoints. The wandering remains Brownian and the correlators $C_D$ and $C_X$ still decay exponentially. The wandering $D(L_y/2)$ for this topologically unconstrained case is plotted in the main panel of Fig.~\ref{Unc wandering} as a function of $L_y$. The results shown are for samples with $L_x=200$ (convergence to the limit $L_x\to\infty$ is  again fast). As expected, the wandering is compatible with Brownian behaviour $D\sim L_y^\nu$ with $\nu = 1/2$ (a fit for $L_y\ge16$ gives exponent $\nu\sim0.51$; the exponent decreases towards $0.5$ when dropping small system sizes). As a further check the inset shows behaviour for $y\ll L_y$, again giving results compatible with Brownian behaviour.

\section{Heuristic approaches}
\label{heuristic arguments appendix}

\begin{figure}[t]
 \begin{center}
\includegraphics[width=0.6\linewidth]{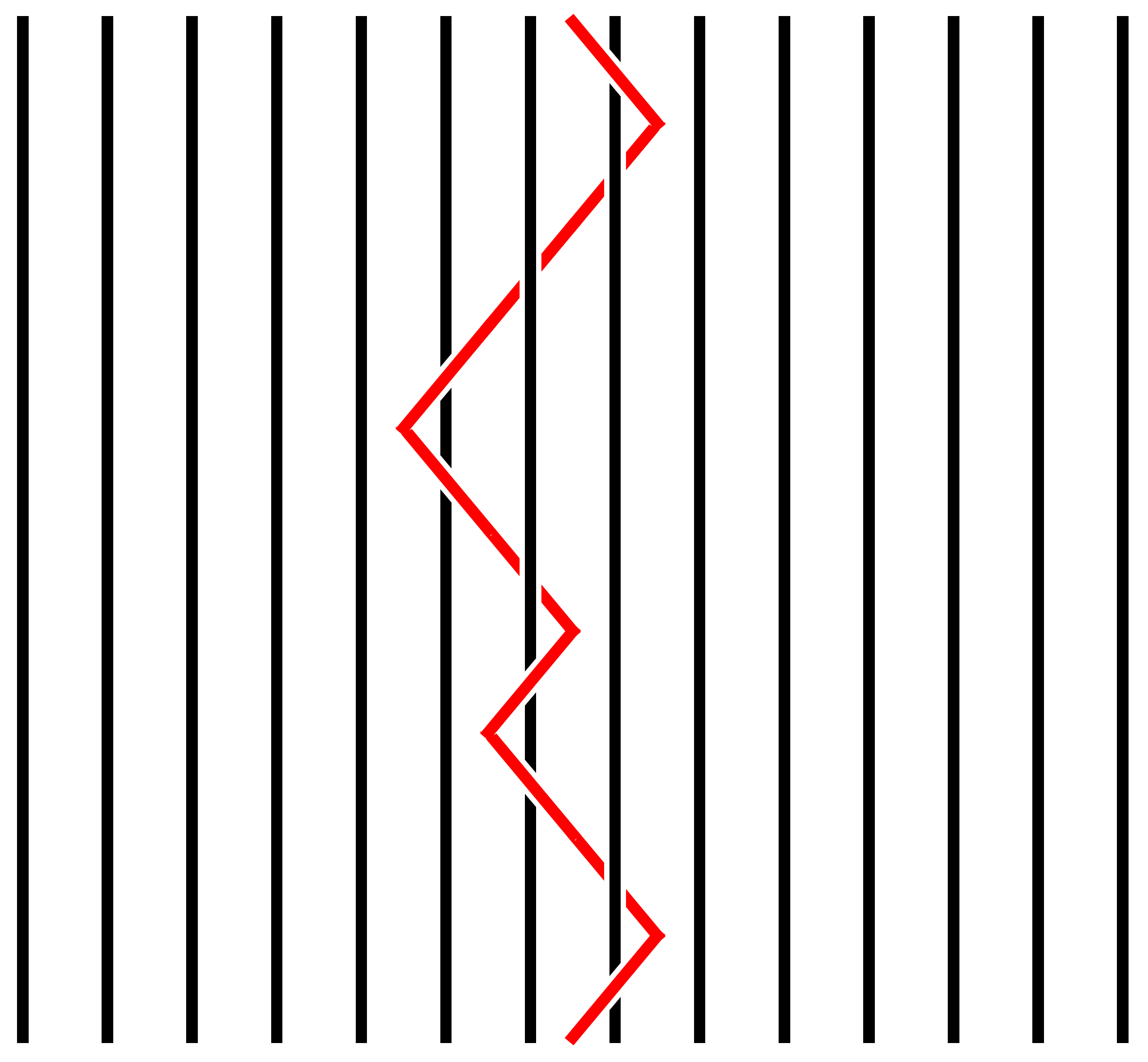}
 \end{center}
 \caption{Toy `array of obstacles' model.}
 \label{toymodel}
\end{figure}

\noindent
\textit{Array of obstacles.} Fig.~\ref{toymodel} illustrates this toy model. There are fixed vertical strands at integer values of the $x$-coordinate, and the mobile strand winds in and out. The endpoints of the mobile strand at $y=0$,  $y=L_y$ are fixed at $x=1/2$, and the global topology is constrained to be trivial (the mobile strand can be deformed to vertical without passing through any background strands). Variants of this model have been considered many times in the literature  \cite{ring_AOO_Nechaev, ring_AOO_Rubinstein, Cates and Deutsch, grest kremer milner witten}. Remarkably, the mathematical structure of this toy model is identical to that of the array of obstacles model for the 3D undirected case, despite the different physical interpretations. The exact exponents quoted in the text ($\nu = 1/4$ and $\mathcal{C}\sim L^{1/2}$) come from an elegant mapping to random walks on the Cayley tree \cite{ring_AOO_Nechaev}. 

\textit{Entropic arguments.} A naive approach to our model is to  minimise a free energy per strand, $F=F_\text{confinement} + F_\text{topological}$, as a function of the typical scale of wandering, $D$. The first term is the cost of confining a free directed strand in a $D$--sized `box': $F_\text{confinement}\sim L_y / D^2$. The second is the free energy cost of not being entangled --- i.e. of of obeying the constraint that the melt is topologically trivial --- and is harder to estimate. The constraint of topological triviality can be regarded as a union of an infinite number of `k-strand' constraints, where by a $k$-strand constraint we mean the requirement that  a given group of $k$ strands is not entangled.

First let us confirm that the \emph{two}-strand constraints alone are far too weak to generate logarithmic confinement (two-strand \textit{constraints} should not be confused with the two-strand \textit{moves} discussed in Sec. II above).  If $d$ is the number of transverse dimensions ($d=1$ in the model simulated), our chosen strand has the opportunity to wind around $O(D^d)$ other strands, each one $O(L_y/D^d)$ times. Viewing $y$ as time, the winding number of a given pair performs a 1D random walk with $O(L_y/D^d)$ steps, which must return to the origin. This gives $F_\text{2-strand} \sim D^d \ln (L_y/D^d)$. Minimising $F$ gives $D\sim L^{\nu}$ (neglecting logarithms) with $\nu = 1/(2+d)$. This confirms that the two-strand constraints are too weak to reproduce what we see: genuinely multi-strand entanglement is crucial. The above result does however support the expectation that wandering is at least as strongly suppressed in 3D as in 2D, as a result of the larger number of other strands encountered for a given typical scale of wandering.

Attempting to go beyond the two strand constraints illustrates the fact that the results will be sensitive to the approximate form assumed for the free energy. We may imagine  successively imposing the 2-strand constraints, then the 3-strand constraints that are independent of them, then the additional 4-strand constraints etc. At each stage we pay a free energy $F_\text{$k$-strand}$ for forbidding configurations with `$k$-partite' entanglement. The crudest approximation is to approximate $F_\text{$k$-strand}$ as a sum of independent terms, one for each group of $k$-strands  which includes the chosen strand and which are not spatially disjoint. (Note that two distinct $k$-strand groups may share some smaller subset of their strands.) This gives $F_\text{$k$-strand}\sim D^{k-1} f_k$ (for large $D$ and $d=1$), where $f_k$ is the free energy cost of forbidding $k$-partite entanglement for a single $k$-strand group. It is tempting to imagine that a conservative approximation is to take $f_k$ to be at least of order one. Then, if we truncate the free energy at some fixed $k$, we obtain an exponent $\nu_k$ which indeed tends to zero as $k\rightarrow \infty$.

This suggests that interactions between arbitrarily large numbers of strands are necessary to explain the observed fact that $\nu=0$. However, the argument may be misleading; treating the groups as independent is an uncontrolled approximation. The array of obstacles model shows the danger. Naively we might attempt a similar argument there ($k$-strand entanglement can arise even if only one strand is mobile) but we know that in that case $\nu$ is not equal to zero. It is an interesting question whether in the full model $k$-strand constraints for arbitrarily large $k$ are necessary to obtain $\nu=0$.

\textit{The renormalisation group.}  One of the most fundamental questions is whether there exists a real-space renormalisation group approach to topologically constrained polymer ensembles (see e.g. Ref.~\cite{redner reynolds} for the case without constraints). Our result that $D$ is only logarithmically large in $L_y$ suggests that the appropriate protocol here would be to rescale the vertical coordinate but not the horizontal one, i.e. not to decimate strands. One must of course find a way to deal with the nonlocality of the topological constraints. One possibility is not to directly coarse-grain the 2D configuration, but instead to focus on the configurations in 2+1D spacetime, with the same time evolution as the Monte Carlo simulations. Spacetime configurations may be viewed in terms of `worldlines' of crossings. The potential simplification is that the global topological constraints are then encoded in \emph{local} constraints on the worldline configurations. If the problem is simplified by dropping the three strand moves from the dynamics (as discussed in the text) we obtain a 2+1D statistical mechanics problem with a concise definition. This is a multi--layer loop model, with oriented loops, in which loops from adjacent layers cannot cross. It is conceivable that this model is tractable by field theory techniques.


\begin{thebibliography}{250}

\bibitem{de gennes reptation 1971} P. G. de Gennes, J. Chem. Phys. {\bf 55}, 572 (1971).
\bibitem{DeGennes_book} P. G. de Gennes, \textit{Scaling Concepts in Polymer Physics}, Cornell University Press, Ithaca (1979).
\bibitem{doi edwards} M.  Doi, and S. F. Edwards, \textit{The Theory of Polymer Dynamics}, Clarendon, Oxford (1986).
\bibitem{Edwards_AOO} S. F. Edwards, British Polym.  J. {\bf 9}, 140 (1977).
\bibitem{ring_AOO_Nechaev} A. R. Khokhlov, and S. K. Nechaev, Phys. Lett. A {\bf 112}, 156 (1985).
\bibitem{ring_AOO_Rubinstein} M. Rubinstein, Phys. Rev. Lett. {\bf 57}, 3023 (1986).
\bibitem{Cates and Deutsch} M.E. Cates and J. M. Deutsch, J. Phys. (Paris) {\bf 47}, 2121 (1986).
\bibitem{Sakaue_Flory} T. Sakaue,  Phys. Rev. Lett. {\bf 106} 167802 (2011); T. Sakaue, Phys. Rev. E {\bf 85}, 021806 (2012).
\bibitem{Grosberg flory arguments} A. Y. Grosberg, Soft Matter {\bf 10}, 560 (2014).
\bibitem{ring_AOO_dynamics} S. P. Obukhov, M. Rubinstein, and T. Duke, Phys. Rev. Lett. {\bf 73}, 1263 (1994).
\bibitem{muller} M. M\"uller, J. P. Wittmer, and M. E. Cates, Phys. Rev. E {\bf 61}, 4078 (2000).
\bibitem{suzuki} J. Suzuki, A. Takano, T. Deguchi, and Y. Matsushita, J. Chem. Phys., {\bf 131}, 144902 (2009).
\bibitem{vettorel} T. Vettorel, A. Y. Grosberg, and K. Kremer, Phys. Biol., {\bf 6}, 025013 (2009).
\bibitem{halverson} J. D. Halverson, G. S. Grest, A. Y. Grosberg, and K. Kremer, Phys. Rev. Lett. {\bf 108}, 038301 (2012).
\bibitem{michieletto} D. Michieletto, D. Marenduzzo, E. Orlandini, G. P. Alexander, and M. S. Turner, ACS Macro Lett. {\bf 3}, 255 (2014).
\bibitem{imakaev et al} M. V. Imakaev, K. M. Tchourine, S. K. Nechaev, and L. A. Mirny, Soft Matter {\bf 11}, 665 (2015).
\bibitem{tamm}  M. V. Tamm, L. I. Nazarov, A. A. Gavrilov, and A. V. Chertovich, Phys. Rev. Lett. {\bf 114}, 178102 (2015).
\bibitem{chromosome_review} J. D. Halverson, et. al., Rep. Prog. Phys. {\bf 77}, 022601 (2014).
\bibitem{melt_viscosity_experiment} M. Kapnistos et. al., Nature Mater. {\bf 7}, 997 (2008).
\bibitem{crumpled globule} A. Y. Grosberg, S. K. Nechaev, and E. I. Shakhnovich, J.  Phys. {\bf 49}, 2095 (1988).
\bibitem{fractal globule} L. A. Mirny, Chromosome Res. {\bf 19}, 37 (2011).
\bibitem{Nechaev_lecture_notes} S. K. Nechaev, \textit{Statistics of Knots and Entangled Random Walks,} World Scientific, Singapore (1996). 
\bibitem{Nechaev_long_braid} S. K. Nechaev, A. Y. Grosberg and A. M. Vershik, J. Phys. A: Math. Gen. {\bf 29}, 2411 (1996).
\bibitem{heap} A.M. Vershik, S. Nechaev and R. Bikbov, Commun. Math. Phys. {\bf 212}, 469 (2000).
\bibitem{B3}  S. Nechaev and R. Voituriez,  J. Phys. A {\bf 36}, 43 (2003).
\bibitem{ferrari review} F. Ferrari, Ann. Phys. {\bf 11}, 255 (2002).
\bibitem{brushes_Alexander} S. Alexander, J. Phys. (Paris) {\bf 38}, 983 (1977).
\bibitem{brushes_deGennes} P. G. De Gennes, Macromolecules {\bf 13}, 1069 (1980).
\bibitem{PBC footnote} Strictly speaking this differs from the standard braid group by the choice of periodic BCs.
\bibitem{Braids_book}  C. Kassel and V. Turaev, \textit{Braid Groups}, Springer-Verlag, New York, 2010.
\bibitem{reith brush dynamics} D. Reith, A. Milchev, P. Virnau, and K. Binder, Macro-
molecules {\bf 45}, 4381 (2012).
\bibitem{power law footnote} Trying $D = A L_y^{\nu} ( 1 + B L_y^{-a})$ gives $\nu=0.10(14)$ and $a=0.1(4)$.
\bibitem{rosa everaers} A. Rosa and R. Everaers, Phys. Rev. Lett. {\bf 112}, 118302 (2014).
\bibitem{grest kremer milner witten} G. S. Grest, K. Kremer, S. T. Milner, and T. A. Witten, Macromolecules {\bf 22}, 1904 (1989).
\bibitem{redner reynolds} S. Redner and P. J. Reynolds, J. Phys. A: Math. Gen. {\bf 14}, 2679 (1981).
\end{thebibliography}
\end{document}